\documentclass[modern, a4paper, times]{aastex63}
\usepackage{amsmath}
\usepackage{graphicx}
\usepackage{natbib}

\submitjournal{ApJL}

\shorttitle{Influence of Solar FUV/EUV Anisotropy on Helioglow}
\shortauthors{Strumik et al.}

\begin{document}

\title{Influence of Heliolatitudinal Anisotropy of Solar FUV/EUV Emissions on Lyman-$\alpha$ Helioglow: SOHO/SWAN Observations and WawHelioGlow Modeling}

\correspondingauthor{M. Strumik}
\email{maro@cbk.waw.pl} 

\author[0000-0003-3484-2970]{M. Strumik}
\affil{Space Research Centre PAS (CBK PAN), Bartycka 18a, 00-716 Warsaw, Poland}

\author[0000-0003-3957-2359]{M. Bzowski}
\affil{Space Research Centre PAS (CBK PAN), Bartycka 18a, 00-716 Warsaw, Poland}

\author[0000-0002-5204-9645]{M. A. Kubiak}
\affil{Space Research Centre PAS (CBK PAN), Bartycka 18a, 00-716 Warsaw, Poland}

\begin{abstract}
Observations of the Sun’s surface suggest a nonuniform radiated flux as related to the presence of bright active regions and darker coronal holes. The variations of the FUV/EUV source radiation can be expected to affect the Lyman-$\alpha$ backscatter glow measured by spaceborne instruments. In particular, inferring the heliolatitudinal structure of the solar wind from helioglow variations in the sky can be quite challenging if the heliolatitudinal structure of the solar FUV/EUV radiation is not properly included in \added{the} modeling of the heliospheric glow.

We present results of analysis of the heliolatitudinal structure of the solar Lyman-$\alpha$ radiation as inferred from comparison of SOHO/SWAN satellite observations of the helioglow intensity with modeling results obtained from \added{the} recently-developed WawHelioGlow model. We find that in addition to time-dependent heliolatitudinal anisotropy of the solar wind, also time-dependent heliolatitudinal variations of the intensity of the solar Lyman-$\alpha$ and photoionizing emissions must be taken into account to reproduce the observed helioglow modulation in the sky. We present a particular latitudinal and temporal dependence of the solar Lyman-$\alpha$ flux obtained as a result of our analysis. We analyze also differences between polar-equatorial anisotropies close to the solar surface and seen by an observer located far from the Sun. We discuss \added{the} implications of these findings for \added{the} interpretation of heliospheric-glow observations.
\end{abstract}

\section{Introduction}
\label{sec:intro}
Many interesting physical phenomena appear as a result of \added{the} interaction of plasma and radiation of solar origin with the interstellar medium. Among others, the heliospheric Lyman-$\alpha$ ($\sim \! \!121.567$ nm) glow provides an opportunity of investigating complex interactions between \deleted{the} interstellar matter, solar wind\added{,} and solar radiation. The Sun is moving relative to the interstellar medium (at $\sim \!\! 26$ km/s; see, e.g., \replaced{\citet{bzowski_etal:08a,bzowski_etal:15a,swaczyna_etal:18a}}{\citet{lallement_etal:04a,witte:04,bzowski_etal:08a,bzowski_etal:15a,schwadron_etal:15a,swaczyna_etal:18a}}) and the heliospheric boundary called the heliopause does not affect significantly neutral H atoms of interstellar origin, which \replaced{are able to}{can} penetrate the heliospheric interface deeply, reaching distances of several AU from the Sun. Very close to the Sun, the distribution of neutral hydrogen is strongly influenced by charge exchange with time- and heliolatitude-dependent solar wind \citep{bzowski:03,sokol_etal:19b}. Also\added{,} FUV/EUV output from the Sun can be expected to affect H atoms by photoionization processes, Lyman-$\alpha$ illumination of H atoms, and by the radiation pressure. In addition to its global time variations, the FUV/EUV output from the Sun may also exhibit a latitudinal dependence \citep{cook_etal:81a,pryor_etal:92,auchere_etal:05a}.
The interaction of H atoms with the solar wind and radiation leads to \added{the} formation of a cavity around the Sun, where the neutral H density is very low. The Lyman-$\alpha$ helioglow is generated by \added{the} scattering of solar photons on H atoms surrounding the cavity.

The helioglow distribution in the sky as seen by an observer located at $\sim\!\!1$ AU or closer to the Sun has been measured by several spaceborne instruments, e.g., Mariner \citep{ajello:78}, Galileo and Pioneer Venus ultraviolet spectrometers \citep{pryor_etal:92}, and the SOHO/SWAN instrument \citep{bertaux_etal:95}. The GLOWS instrument onboard a future IMAP mission \citep{mccomas_etal:18b} will be focused on measuring the helioglow distribution in the sky and inferring the heliolatitudinal structure of the solar wind based on the measurements and modeling the helioglow.

In terms of modeling, there has been \deleted{a} significant progress in understanding the properties of the backscatter glow and associated physical mechanisms. Based on early papers by Hummer \citep{hummer:62a,hummer:64a, hummer:68a, hummer:69a, hummer:69b}, various modeling approaches have been proposed for the backscatter glow \citep{weller_meier:74, meier:77a, keller_thomas:79a, keller_etal:81a, quemerais_bertaux:93a, scherer_fahr:96, quemerais:00, quemerais:06a, fayock_etal:13a}. A recently developed WawHelioGlow model \citep{kubiak_etal:21a,kubiak_etal:21b} incorporates fully kinetic treatment of multiple populations of neutral H atoms. The model also includes a realistic temporal dependence and latitudinal anisotropy of the solar wind parameters, which together with an observation-based model of FUV/EUV solar output determine the distribution of the H atoms in configuration and velocity space in the proximity of the Sun.

In this Letter, we investigate the influence of the heliolatitudinal structure of the solar FUV/EUV output on the Lyman-$\alpha$ backscatter helioglow. For this purpose\added{,} we use the WawHelioGlow model, where the latitudinal structure of solar wind is included as retrieved from interplanetary scintillations of compact radio sources \citep{tokumaru_etal:10a,tokumaru_etal:12b,sokol_etal:20a}. We show that by including additionally the anisotropic FUV/EUV structure, the WawHelioGlow model provides \added{a} much better fit to SOHO/SWAN observations of the backscatter glow in the solar maximum as compared with isotropic-radiation modeling. We investigate different predefined anisotropy levels and compare \added{the} resulting helioglow distribution in the sky with SOHO/SWAN observations.

The Letter is organized as follows. In Section \ref{sec:model} we discuss a solar-FUV/EUV-output anisotropy relation adopted in the WawHelioGlow model. Examples of modeled helioglow sky distributions and their comparison with SOHO/SWAN observations are discussed in Section \ref{sec:modeling_smin_smax}. Time dependence of the anisotropy inferred from SOHO/SWAN observations and WawHelioGlow modeling is shown in Section \ref{sec:anis_time_dep}. A heliodistance dependence of the anisotropy is discussed in Section \ref{sec:anis_dist_dep}. Our conclusions are summarized in Section \ref{sec:discussion}.

\section{Modeling Anisotropy of Solar FUV/EUV Output}
\label{sec:model}
For modeling the helioglow distribution in the sky we use a recently developed WawHelioGlow model of the helioglow \citep{kubiak_etal:21a,kubiak_etal:21b}. In the model, the helioglow flux density in the sky is calculated using an optically thin, single-scattering approximation. The distribution of H atoms around the Sun is calculated from the Warsaw Test Particle Model (n)WTMP \citep{tarnopolski_bzowski:09}, where \added{the} fully kinetic treatment of the distribution function for H atoms is applied. Solar-wind temporal and latitudinal variations are included as inferred from interplanetary scintillations (IPS) and in-ecliptic measurements provided in the OMNI database \citep{bzowski_etal:13a,sokol_etal:20a}. Solar radiation pressure, which modifies trajectories of H atoms in the heliosphere is modeled following \citet{IKL:20a}.

Apart from the temporal and latitudinal structure of the solar wind, the WawHelioGlow model takes into account the latitudinal anisotropy of the solar FUV/EUV output. The entire FUV/EUV flux (i.e., both the illuminating and ionizing parts) is modeled as modulated with the heliographic latitude $\phi$
\begin{equation}
E=E_\mathrm{eq} [a \sin^2\phi +\cos^2\phi].
\label{eq:anis}
\end{equation}
The parameter $a=E_\mathrm{p}/E_\mathrm{eq}$ defines the ratio of the polar irradiance $E_\mathrm{p}$ to the equatorial irradiance $E_\mathrm{eq}$. It is important to note that the irradiance in Equation (\ref{eq:anis}) should be understood as defined at the location of \replaced{a}{an} H atom that either scatters or absorbs \replaced{a}{an} FUV/EUV photon. Therefore, in general\added{,} the irradiance anisotropy in Equation (\ref{eq:anis}) can be significantly different from the anisotropy of the solar-surface radiance derived from synoptic maps. This question is discussed in a more detail in Section \ref{sec:anis_dist_dep}. The adopted model is symmetric in $\phi$, i.e., the same irradiance is set for the north and south pole $E_\mathrm{p}=E(\phi=\pi/2)=E(\phi=-\pi/2)$. It was shown by \citet{kubiak_etal:21a} that the anisotropy model of Equation (\ref{eq:anis}) is fully equivalent to other approach proposed by \citet{pryor_etal:92}. In principle, latitude-resolved observations of the Sun (e.g., synoptic FUV/EUV maps of the solar surface) could provide time-dependent anisotropy input data for the WawHelioGlow model, but for the Lyman-$\alpha$ wavelength\added{,} such observations are currently lacking.

The adopted model of heliolatitudinal anisotropy of the solar FUV/EUV output affects (1) the radiation pressure acting on H atoms, (2) the photoionization rate of the atoms, and (3) the illumination of the atoms in the process of helioglow formation.
\section{Helioglow-modeling Results for Solar Minimum and Maximum}
\label{sec:modeling_smin_smax}

\begin{figure}[!htbp]
\begin{center}
\includegraphics[scale=0.42]{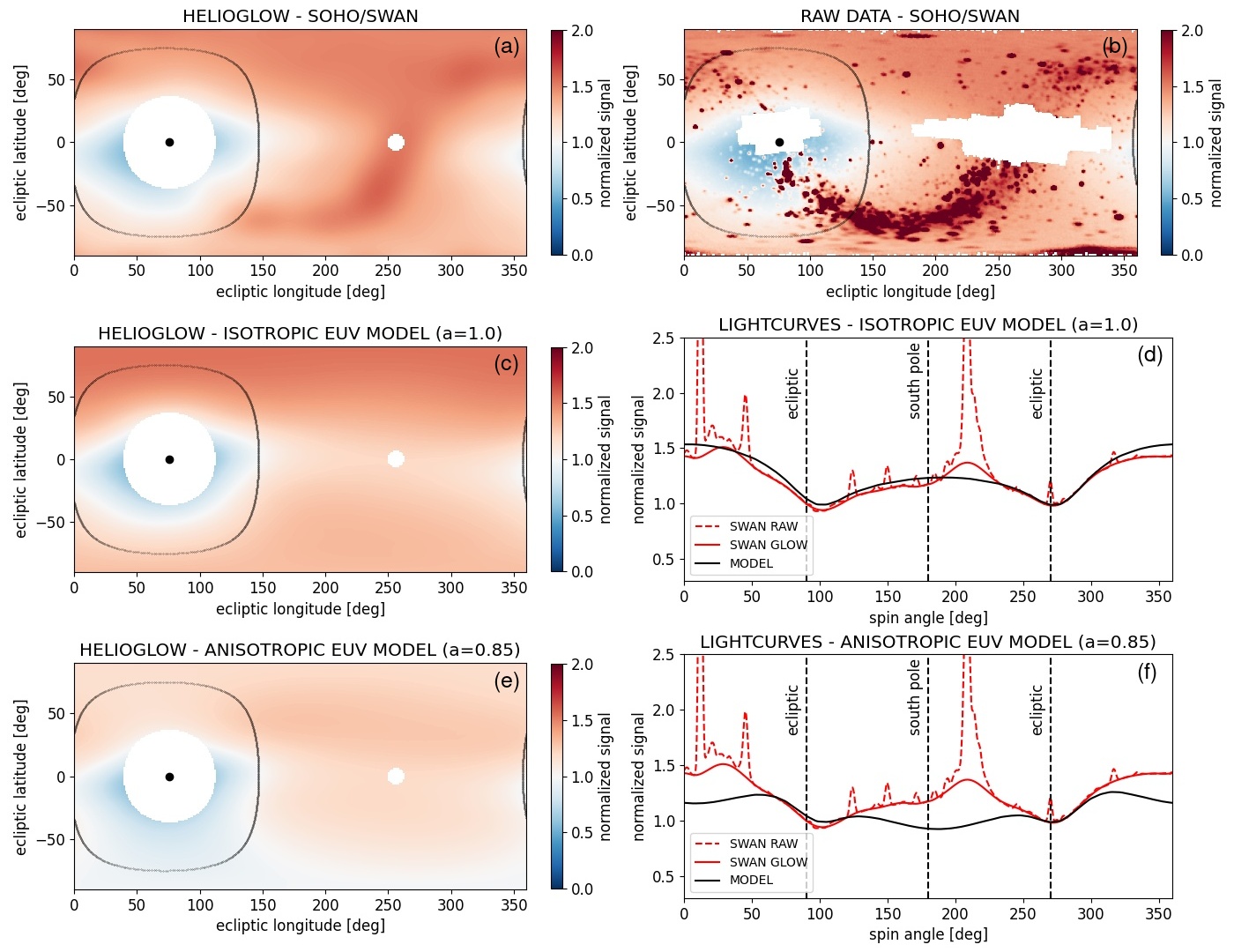}
\caption{Comparison of SOHO/SWAN observations with WawHelioGlow modeling results for the upwind location of the observer and solar minimum conditions (June 5, 1996). An observed helioglow map (panel (a)) is extracted from a raw SOHO/SWAN map (panel (b)) by filtering out extra-heliospheric point sources. Simulated helioglow maps are shown for the isotropic solar FUV/EUV output (panel (c), $a=1$) and an anisotropic case (panel (c), $a=0.85$). The ecliptic coordinates are used in the maps. \replaced{Black}{The black} dot in the maps shows the Sun position, while white regions are masked out due to their proximity to solar and anti-solar directions. The black circular path in the maps (a)-(c) and (e) shows a scanning circle (angular radius 75 deg) for the planned GLOWS instrument onboard the IMAP satellite. Panels (d) and (f) present a comparison of signal modulation along the scanning circle for the observed and modeled helioglow. Spin angle is measured \replaced{counter clockwise}{counterclockwise} from the northernmost location.}
\label{fig:swan_vs_model_smin}
\end{center}
\end{figure}

Figure \ref{fig:swan_vs_model_smin} shows a comparison of selected SOHO/SWAN observations and WawHelioGlow modeling results. Raw SOHO/SWAN observations obtained on June 5, 1996 (solar minimum conditions) are presented in Figure \ref{fig:swan_vs_model_smin}(b). At this date, the satellite was located approximately at the upwind direction relative to the local interstellar medium (LISM) flow. The map displays contributions both from extra-heliospheric point sources and the helioglow slowly varying in the sky. To remove the point-source contributions, we use a procedure described in Section 4.1 of \citet{strumik_etal:20a}. This procedure consists in using machine-learning-based techniques to find a smooth approximation of large-scale features in the SOHO/SWAN maps that are interpreted as the helioglow. The extracted helioglow is shown in Figure \ref{fig:swan_vs_model_smin}(a). Figures \ref{fig:swan_vs_model_smin}(c) and (e) show sky maps with results of WawHelioGlow modeling for $a=1$ (isotropic solar FUV/EUV output) and $a=0.85$ (anisotropic model), correspondingly. In all three cases (i.e., Figures \ref{fig:swan_vs_model_smin}(a), (c), and (e)), the same signal normalization procedure is applied, where the observed/modeled signal is divided by the average value over an ecliptic equatorial belt ($\pm 0.5$ deg from the equator, as determined by the grid on which the SOHO/SWAN maps are provided). We assume that the normalized signal from observations and our modeling results can be compared directly, as insensitive to possible absolute-calibration uncertainties.

The modeled helioglow maps show general similarity to the SOHO/SWAN observations. The map shown in Figure \ref{fig:swan_vs_model_smin}(c) is superior relative to Figure \ref{fig:swan_vs_model_smin}(e), due to a better agreement with Figure \ref{fig:swan_vs_model_smin}(a) at mid and high ecliptic latitudes. This conclusion is confirmed by comparison of \replaced{lightcurves}{light curves} shown in Figures \ref{fig:swan_vs_model_smin}(d) and (f), which describe modulation of the signal along a scanning circle of angular radius of 75 deg centered approximately at the Sun position. All these results suggest that for the solar minimum the isotropic model of the solar FUV/EUV output works better than the anisotropic one.

\begin{figure}[!htbp]
\begin{center}
\includegraphics[scale=0.42]{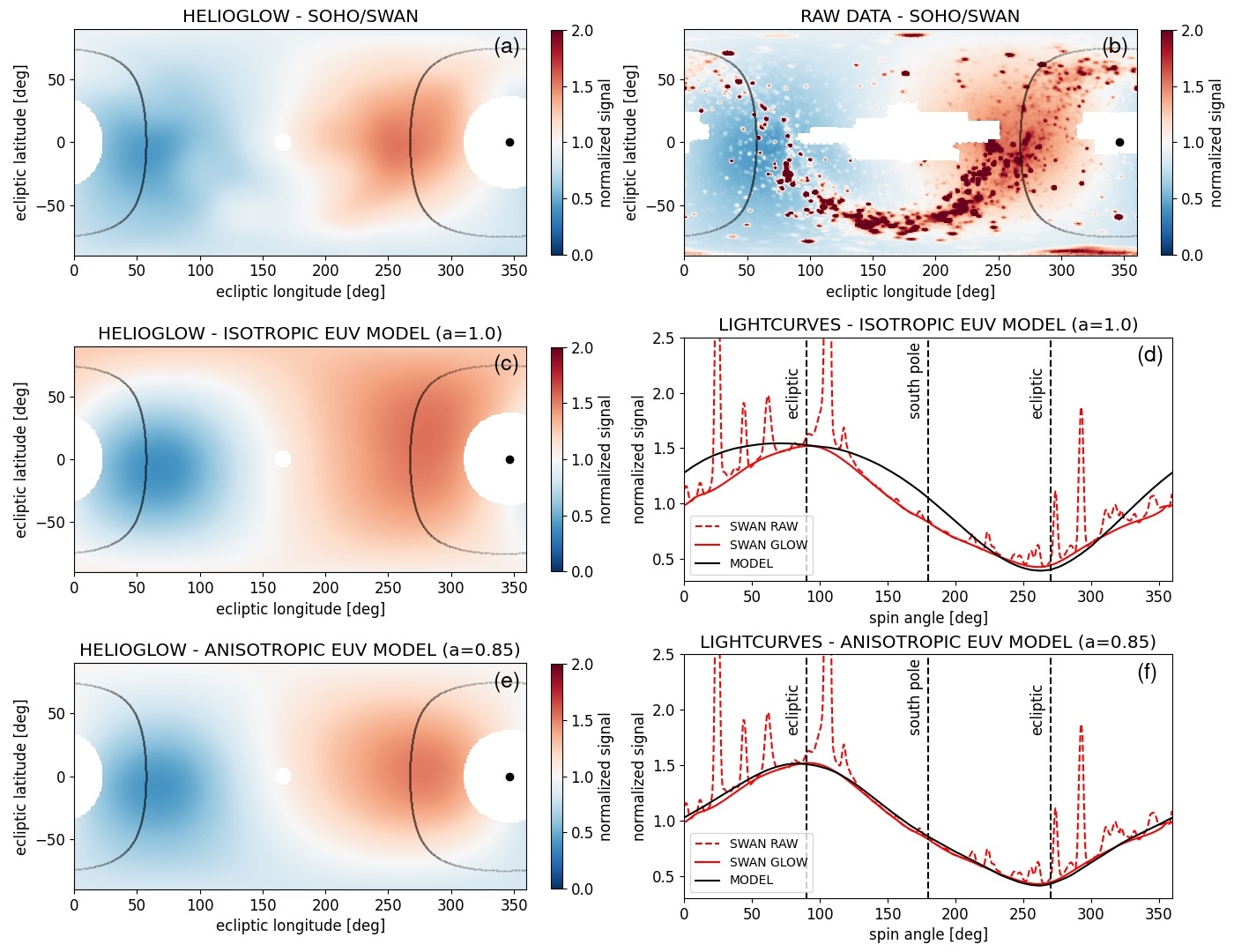}
\caption{A similar comparison of SOHO/SWAN observations with WawHelioGlow modeling results as in Figure \ref{fig:swan_vs_model_smin} but for the crosswind location of the observer and solar maximum conditions (March 6, 2001).}
\label{fig:swan_vs_model_smax}
\end{center}
\end{figure}

A similar comparison but for the crosswind location of the observer and solar maximum conditions (March 6, 2001) is shown in Figure \ref{fig:swan_vs_model_smax}. Again, WawHelioGlow maps in Figures \ref{fig:swan_vs_model_smax}(c) and (e) reproduce general features of the observed distribution of helioglow flux in the sky shown in Figure \ref{fig:swan_vs_model_smax}(a). However, now the anisotropic FUV/EUV model (Figure \ref{fig:swan_vs_model_smax}(e)) provides a much better fit to observations as compared to isotropic modeling. The isotropic model (Figure \ref{fig:swan_vs_model_smax}(c)) apparently yields excessive helioglow intensity at high latitudes, especially in the north. This effect is clearly seen also in lightcurves shown in Figures \ref{fig:swan_vs_model_smax}(d) and (f). By including additionally the anisotropic FUV/EUV output we obtain a signal decrease at exactly those spin-angle sectors, where the isotropic model displays discrepancies with respect to the SOHO/SWAN observations. It is important to emphasize that solar-wind temporal and latitudinal variations have been also included in the model as inferred from IPS and OMNI data \citep{sokol_etal:20a}.

\begin{figure}[!htbp]
\begin{center}
\includegraphics[scale=0.45]{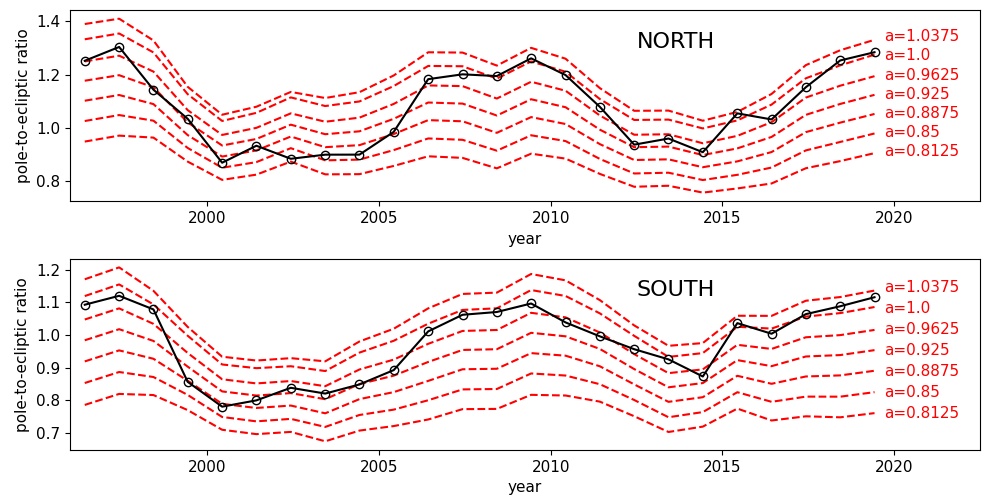}
\caption{Comparison of polar-to-ecliptic ratios of the helioglow signal for SOHO/SWAN observations (black)\replaced{, isotropic FUV/EUV model ($a=1$, red) and anisotropic ($a=0.85$, green) case}{and the WawHelioGlow model (red) with different anisotropy levels defined by the parameter $a$}. The upper panel shows the results for the north pole and the lower panel for the south pole.}
\label{fig:polar_eclipt_ratio}
\end{center}
\end{figure}

\section{Inferred Time Dependence of FUV/EUV Anisotropy}
\label{sec:anis_time_dep}
The inclusion of the polar anisotropy of the solar FUV/EUV output naturally largely affects the polar-to-ecliptic ratio of the helioglow intensity. Therefore, it is interesting to see how the ratio varies in time. Figure \ref{fig:polar_eclipt_ratio} shows a comparison of the polar-to-ecliptic ratio for SOHO/SWAN observations as compared with the WawHelioGlow model \replaced{both the isotropic ($a=1$) and anisotropic ($a=0.85$) cases}{for different levels of solar FUV/EUV output anisotropy from $a=0.8125$ to $a=1.0375$}. The relations for the north and south poles are shown separately for the years 1996-2020. The polar signals are averaged over the northern and southern polar caps of the angular radius of 5 deg. The equatorial signal is averaged over a $\pm 10$-deg belt around the ecliptic equator. For \added{the} SOHO/SWAN \added{observations}, both the polar and equatorial signals have been computed after cleaning the maps from point-source (presumably stellar) \replaced{contamination's}{contaminations}.

The general pattern revealed by Figure \ref{fig:polar_eclipt_ratio} can be described as follows. The observed polar-to-ecliptic ratio (black curve) changes in time but remains \replaced{well}{approximately} within the boundaries established by the WawHelioGlow model with the isotropic solar FUV/EUV output ($a=1$\deleted{, red}) and \added{the} anisotropic \replaced{($a=0.85$, green curve)}{$a=0.85$} case. For the periods of solar-minimum conditions ($\sim \!\!1996$, $\sim \!\!2009$, $\sim \!\!2019$), generally the isotropic model shows a better agreement with observations. On the other hand, for the periods of solar-maximum conditions ($\sim \!\!2002$, $\sim \!\!2014$), a significant anisotropy of the solar FUV/EUV output is required to reproduce the SOHO/SWAN observations.

\replaced{Based on the polar-to-ecliptic ratios shown in Figure \ref{fig:polar_eclipt_ratio}, we can estimate the expected time evolution of the solar-FUV/EUV-output anisotropy $A$ using the formula
\begin{equation}
A=0.85+0.15~\frac{R_\mathrm{obs}-R_\mathrm{a=0.85}}{R_\mathrm{a=1}-R_\mathrm{a=0.85}}.
\label{eq:anis_inferred}
\end{equation}
Here, $R_\mathrm{obs}$ denotes the observed SOHO/SWAN ratio (black curve in Figure \ref{fig:polar_eclipt_ratio}), $R_\mathrm{a=1}$ is the ratio obtained for the isotropic model (red curve in Figure \ref{fig:polar_eclipt_ratio}) and $R_\mathrm{a=0.85}$ is the ratio for the anisotropic model (green curve in Figure \ref{fig:polar_eclipt_ratio}).}
{The quantity that can be directly determined from Figure \ref{fig:polar_eclipt_ratio} is the ratio $R_\mathrm{obs}$ for the SOHO/SWAN observations (black circles) for each date. Figure \ref{fig:polar_eclipt_ratio} also shows that the modeled dependence of the ratio $R(a)$ is monotonic for a fixed date. Therefore, we can define the inverse function $a(R)$ and use the piecewise linear interpolation of the modeled dependence to compute the inferred anisotropy $A=a(R_\mathrm{obs})$. Repeating this procedure for each date presented in Figure \ref{fig:polar_eclipt_ratio}, we can estimate the expected time evolution of the solar-FUV/EUV-output anisotropy.}

\begin{figure}[!htbp]
\begin{center}
\includegraphics[scale=0.4]{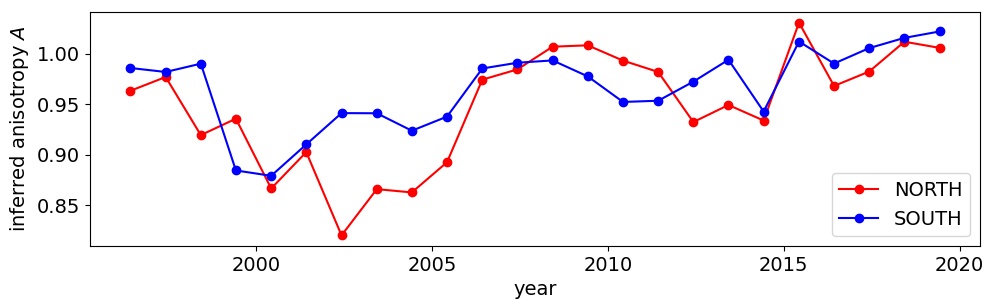}
\caption{Inferred time dependence of the anisotropy $A$ of the solar FUV/EUV output far from the Sun at locations, where the radiation is scattered or absorbed. Dependencies for the north (red) and south (blue) poles are shown.}
\label{fig:inferred_anis}
\end{center}
\end{figure}

Figure \ref{fig:inferred_anis} shows the estimated anisotropy of the solar FUV/EUV output for the years 1996-2020 as computed \replaced{from Equation (\ref{eq:anis_inferred})}{using the procedure described above}. For the solar maximum of $\sim\!\!2002$, the estimated anisotropy level is $7-18\%$, while for the solar maximum of $\sim\!\!2014$ it is significantly smaller $\sim\!\!5\%$. North-south differences of the anisotropy are also clearly visible in the plot.

\section{Heliodistance Dependence of Solar FUV/EUV Irradiance Anisotropy}
\label{sec:anis_dist_dep}
It is important to note that the FUV/EUV irradiance anisotropy discussed so far is defined at the location of \replaced{a}{an} H atom that either scatters \replaced{a}{an} FUV/EUV photon in the process of generation of Lyman-$\alpha$ helioglow or absorbs the photon during the ionization process. Generally, for anisotropic emission from a sphere, the latitudinal profile of irradiance depends on the distance of an observer from the Sun. Having an observer located at $\mathbf{r}_\mathrm{d}$, the irradiance from a spherical source of radius $R_\odot$ can be computed from the following formula
\begin{equation}
E(\mathbf{r}_\mathrm{d})=\int_{\Omega_\mathrm{S}:~ \cos{\alpha}~>0} \frac{L(\mathbf{r}_\mathrm{S})}{(r/R_\odot)^2}~ \cos{\alpha}~ \cos{\omega} ~\mathrm{d}\Omega_\mathrm{S}.
\label{eq:irrad_anis}
\end{equation}
We assume here that the sphere is located at the center of our frame, the position of an element of the sphere $\mathrm{d}\Omega_\mathrm{S}$ is given by a vector $\mathbf{r}_\mathrm{S}$, the vector between the element of the sphere and the observer is $\mathbf{r}=\mathbf{r}_\mathrm{d}-\mathbf{r}_\mathrm{S}$ and $r=|\mathbf{r}|$. The angles in Equation (\ref{eq:irrad_anis}) are defined as follows $\alpha=\arccos(\mathbf{r}_\mathrm{S}\cdot\mathbf{r}/(|\mathbf{r}_\mathrm{S}||\mathbf{r}|))$ and $\omega=\arccos(\mathbf{r}_\mathrm{d}\cdot\mathbf{r}/(|\mathbf{r}_\mathrm{d}||\mathbf{r}|))$. By $L(\mathbf{r}_\mathrm{S})$ we denote the anisotropic radiance that depends on the location on the source sphere.

\begin{figure}[!htbp]
\begin{center}
\includegraphics[scale=0.5]{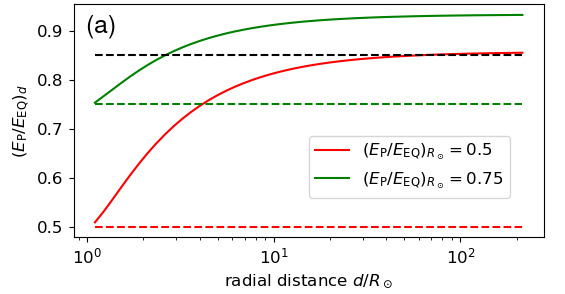}
\includegraphics[scale=0.45]{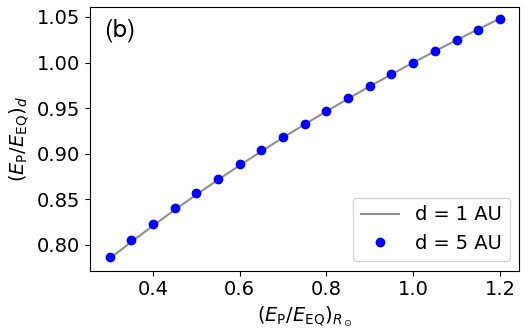}
\caption{Dependencies of the polar-to-equatorial ratio $(E_\mathrm{p}/E_\mathrm{eq})_\mathrm{d}$ on (a) the radial distance $d$ from the Sun and (b) the ratio $(E_\mathrm{p}/E_\mathrm{eq})_\mathrm{R_\odot}$ measured at the solar surface. In panel (a) dependencies for two different solar-surface ratios $(E_\mathrm{p}/E_\mathrm{eq})_\mathrm{R_\odot}$ are shown: 0.5 (red solid line) and 0.75 (green solid line). The radial distance $d$ is measured in the solar radius $R_\odot$ units. The red and green dashed lines represent solar-surface-anisotropy ratios. The black dashed line shows the level 0.85, which corresponds to WawHelioGlow-modeling setting presented in Section \ref{sec:modeling_smin_smax}. Panel (b) shows computation results for two distances $d$ from the Sun: 1 and 5 AU.}
\label{fig:radiance_ratio}
\end{center}
\end{figure}

Using Equation (\ref{eq:irrad_anis}) and applying the anisotropy model of Equation (\ref{eq:anis}) to the radiance $L(\mathbf{r}_\mathrm{S})$ we can compute numerically the dependence of the polar-to-equatorial irradiance ratio on the distance $d=|\mathbf{r}_\mathrm{d}|$ between an observer and the Sun. The results are shown in Figure \ref{fig:radiance_ratio}(a) for two solar-surface anisotropies: 0.5 and 0.75. Computations are done for $1<d/R_\odot<215$, where the upper limit corresponds to $\sim \!\! 1$ AU. A gradual increase of the ratio $E_\mathrm{p}/E_\mathrm{eq}$ (i.e., anisotropy decrease) is seen when we increase the distance and at large distances $d/R_\odot>100$ the curve levels out. The solar-surface anisotropy set to 0.5 gives the anisotropy level of $\sim \!\!0.85$ at large distances from the Sun, which corresponds to our WawHelioGlow-modeling setting presented in Section \ref{sec:modeling_smin_smax}. \replaced{General}{The general} character of the radial dependence of the anisotropy can be easily understood if we realize that in our model an observer located close to the Sun receives photons that are emitted from a relatively small solar surface element located underneath. On the other hand, an observer located far from the Sun receives photons that are emitted from almost \added{the} entire hemisphere thus contributions from different-radiance sectors are largely averaged.

The results presented above suggest that the solar flux anisotropy at the solar surface (e.g., in \replaced{analysis}{analyses} of synoptic maps \citep{auchere_etal:05a}) is generally different from anisotropy seen by an observer located far from the Sun. A relationship between the two anisotropies is shown in Figure \ref{fig:radiance_ratio}(b). There is no significant change in the relation between 1 and 5 AU, which implies that the irradiance anisotropy for H atoms located beyond 1 AU is a function of the heliolatitude but it does not depend significantly on the distance from the Sun. The relation presented in Figure \ref{fig:radiance_ratio}(b) shows that the solar-surface polar darkening of the order of $\sim\!\!50\%$ corresponds to $\sim\!\!15\%$ at large distances from the Sun. If we consider the irradiance itself at large distances from the Sun, it obviously decreases as $E(d)\propto d^{-2}$.

The solar irradiance anisotropy in the WawHelioGlow model should be understood as defined at the location of the photon scattering/absorption by H atoms. As discussed by \citet{kubiak_etal:21b}, the scattering and absorption typically occur at distances of the order of several AU or more from the Sun. The solar FUV/EUV irradiance anisotropy in the model is assumed to be independent of the heliospheric distance. The results presented in this section generally confirm the validity of this approach.

\section{Discussion and Conclusions}
\label{sec:discussion}
In the WawHelioGlow model, the heliolatitudinal structure of the solar wind is retrieved from interplanetary scintillations, which is an independent source of information with respect to helioglow observations. Unlike other approaches (see, e.g., \citet{katushkina_etal:19a}), the solar-wind heliolatitudinal anisotropy is set independently here, thus it is possible to investigate separately the effects of the anisotropy of the solar FUV/EUV output on the Lyman-$\alpha$ backscatter helioglow. Results presented in Sections \ref{sec:modeling_smin_smax} and \ref{sec:anis_time_dep} clearly suggest that including the heliolatitudinal dependence of the solar FUV/EUV output is important to obtain the helioglow modulation in the sky that is consistent with SOHO/SWAN observations in the solar maximum. Comparison of the observed and modeled lightcurves shows that the FUV/EUV anisotropy corrections compensate properly the helioglow intensity exactly at those spin-angle sectors, where the isotropic model displays discrepancies with respect to the SOHO/SWAN observations. Therefore, our results suggest that the solar FUV/EUV anisotropy effects should be necessarily taken into account in modeling the Lyman-$\alpha$ backscatter helioglow. Since solar FUV/EUV output is important for modeling the neutral hydrogen distribution in the vicinity of the Sun, we can presumably extrapolate this conclusion as involving general aspects of \added{the} modeling of neutrals.

More detailed analysis of time dependence of the solar FUV/EUV anisotropy inferred from comparison of WawHelioGlow modeling with SOHO/SWAN observations shows that the isotropic-flux approximation may be sufficient for some short periods of time around the solar minimum. However, in general\added{,} anisotropy corrections need to be included in modeling to properly capture the helioglow evolution in polar regions. Our analysis suggests different inferred-anisotropy levels for the maxima of solar cycles 23 and 24 that are covered by our analysis. Generally\added{,} the inferred anisotropy levels $5-15\%$ are consistent with results of \citet{cook_etal:81a} analysis.

It is important to realize that the adopted one-parameter model of the solar FUV/EUV anisotropy is arbitrary and most likely represents a strongly simplified approach. \deleted{Therefore our quantitative findings are only approximate.} \replaced{Because of}{However, due to} \added{the} unavailability of actual measurements of the solar FUV/EUV output variability at the solar surface for Lyman-$\alpha$, analysis of helioglow observations must rely on very approximate models. \replaced{However, the}{The} discussion presented in Section \ref{sec:modeling_smin_smax} suggests that even very approximate models can be successfully used in the analysis and yield better results than analyses, where the FUV/EUV anisotropy is neglected altogether.

Using a simple model of solar-radiance anisotropy we show a relation between the anisotropy at the solar surface and the anisotropy seen at large distances from the Sun, in regions where \added{the} Lyman-$\alpha$ backscatter helioglow is generated. Using \replaced{the}{this} relation, for solar cycle 23 we obtain the anisotropy level of the order of $20-50\%$\replaced{that}{, which} is consistent with the solar-surface anisotropy obtained from synoptic maps of the Sun for 30.4 nm by \citet{auchere_etal:05a} for similar dates.

All the results presented in this Letter suggest a consistent picture of the influence of solar FUV/EUV anisotropy effects on the Lyman-$\alpha$ backscatter helioglow modulation in the sky. \replaced{We hope that our results help in understanding different contributions to the helioglow generation process, in particular the distribution of neutral H atoms around the Sun, which is important for spaceborne missions focused on neutrals-based imaging of the heliosphere.}{The results of our work can be significant for analyzes of helioglow observations from, e.g., the SOHO/SWAN instrument \citep{bertaux_etal:95} or the ALICE experiment on New Horizons \citep{gladstone_etal:18a}. The anisotropy effects are particularly important for the interpretation of future IMAP/GLOWS observations \citep{mccomas_etal:18b}, where the latitudinal structure of the solar wind will be investigated based on the helioglow modulation in the sky. This latitudinal structure determines the charge-exchange losses of neutral atoms in the heliosphere, thus the presented results are also relevant for general aspects of the modeling of neutral hydrogen atoms in the heliosphere. In this context, the detailed understanding of other heliolatitude-dependent factors modulating the helioglow, like solar FUV/EUV output, is important.
The solar FUV/EUV anisotropy may be particularly significant for NIS He, which is ionized closer to the Sun as compared to H. For $d<1$ AU, a significant dependence of the effective photoionization rate on both the heliolatitude and heliodistance is expected as suggested by our analysis. Close to the Sun, departures from inverse-square scaling of the irradiance and the ionization rate may appear, which likely affects the He distribution in the tail and the helium helioglow. Therefore, our results can be also expected to be important for both the He (58.4 nm) helioglow and the process of He ionization itself.
In a broader astrophysical context, our results can be potentially interesting for understanding the FUV/EUV variations of Sun-like stars as possibly related to the time-varying spherical-anisotropy effects in addition to well-known global variations related to cycles of stellar activity.}

\acknowledgments{This study was supported by \added{the} Polish National Science Center grant 2019/35/B/ST9/01241 and by \added{the} Polish Ministry for Education and Science under contract MEiN/2021/2/DIR.}

\software{scikit-learn \citep{pedregosa_etal:11}, astropy \citep{astropy:13,astropy:18}}


\end{document}